\documentclass[showpacs,aps,pra,twocolumn]{revtex4}
\usepackage{graphicx,epsfig}
\usepackage{times}
\usepackage{amsmath,amssymb,wasysym}

\newlength\figurewidth
\newcommand{\limacon}{{lima\c con}}

\newcommand{\quoting}[1]{``#1''}

\newcommand{\fig}{Fig.}
\newcommand{\figs}{Figs.}

\newcommand{\rem}[1]{}

\begin{document}
\title{Combining directional light output and ultralow loss in deformed microdisks} 
  \author{Jan Wiersig}
  \affiliation{Institut f{\"u}r Theoretische Physik, Universit{\"a}t Bremen,
  Postfach 330 440, D-28334 Bremen, Germany}
  \author{Martina Hentschel}
\affiliation{Max-Planck-Institut f\"ur Physik komplexer Systeme, N{\"o}thnitzer Str. 38, D-01187
Dresden, Germany}
\date{\today}
\begin{abstract}
A drawback of high-quality modes in optical microdisks is their isotropic
light emission characteristics. Here we report a novel, robust, and general
mechanism that results in highly directional light emission from those modes. 
This surprising finding is explained by a combination of wave phenomena 
(wave localization along unstable periodic ray trajectories) 
and chaotic ray dynamics in open systems (escape along unstable
manifolds). The  emission properties originating in the chaotic ray
dynamics 
permit directional light output even from microlasers operating in the common 
multi-mode regime. We demonstrate our novel mechanism for the {\limacon}
billiard family and find directional emission with narrow angular
divergence for a significant range of geometries and material parameters. 
\end{abstract}
\maketitle

The possibility to confine photons in all three spatial dimensions  using
microcavities has triggered intense basic and applied research in physics over
the past decade~\cite{Vahala03}, e.g., research on ultralow threshold
lasing~\cite{Park04,Ulrich06},
single-photon emitters~\cite{Michler2000}, solid-state cavity quantum
electrodynamics~\cite{RSL04,PSMLHGB05,Schwab06}.
Prominent examples of optical microcavities are whispering-gallery
cavities such as microdisks~\cite{MLSGPL92,MSJPLHKH07},
microspheres~\cite{CLBRH93,GPI00,GMM06}, and microtori~\cite{IGYM01,SIM04} which
trap photons for a long time $\tau$ near the boundary by total internal
reflection. The corresponding
whispering-gallery modes have very high quality factors $Q =
\omega\tau$, where $\omega$ is the resonance frequency. For microdisks, the 
record $Q$-factor is around $6.6\times 10^5$~\cite{MSJPLHKH07}.
The high $Q$-factors and the in-plane light emission make microdisks
attractive candidates for several optoelectronic devices, especially for the
nitride material system~\cite{THSLNH07} where other cavity designs such as
vertical-cavity surface-emitting laser (VCSEL) micropillars faces severe
challenges in mirror fabrications~\cite{LSG05}. Unfortunately, the
possible use of microdisks is limited by the fact that the in-plane light
emission is isotropic.

Shortly after the first fabrication of semiconductor microdisk resonators 
it has been
demonstrated that deforming the boundary of a disk allows for improved
directionality of emission and therefore for more efficient extraction and
collection of light~\cite{LSMGPL93,ND97,GCNNSFSC98,THFH07}. 
Several shapes have been proposed and realized since then, but
only few lead to light emission into a single direction with reasonable angular
divergence~\cite{KLRK04,KTMJCC04} that is 
essential for applications like single-photon sources. 
Moreover, all deformed microdisks discussed in the literature have a
serious problem, $Q$ spoiling~\cite{Noeckel94}: 
The $Q$-factor degrades dramatically upon deformation; in the worst
case ruling out any application. 
The trade-off
between $Q$-factor and directionality is not only a problem of microdisks but
also of microspheres, microtori, and even of VCSEL-micropillars: for given
total number of Bragg mirror pairs such a cavity can be optimized either
w.r.t. directionality (significantly more mirror pairs at the bottom than 
at the top) or $Q$-factor (roughly equal number of mirror pairs at the bottom
and top). 

Recently, a scheme to achieve highly directional emission without $Q$-spoiling has
been suggested~\cite{WH06}. It employs the characteristic modifications of
spatial mode structures near avoided resonance crossings.
The drawback of this approach is the existence of nearly degenerate modes
which can have different far-field patterns (FFPs). If two such modes are
excited simultaneously (typical for present-day devices), the
directionality might be lost. Nearly degenerate modes with similar FFP do
occur but to find them requires laborious numerical calculations and a
sophisticated adjustment of geometry parameters depending on refractive index, 
wavelength and cavity size. 

In this Letter we introduce a novel, robust, and generally applicable 
mechanism to achieve highly directional light emission from high-$Q$ modes 
in microdisks which does not suffer from the above-mentioned drawback. 
The key to our finding is to apply what is at
the heart of quantum chaos~\cite{Stoeckmann00} and nonlinear
dynamics~\cite{LichLieb92} of open systems, 
respectively: wave localization along unstable periodic ray
trajectories in
systems with chaotic ray dynamics~\cite{Heller84} and the only recently
acknowledged importance of the so-called unstable manifold for the FFPs of
microcavities~\cite{SRTCS04,LRKRK05,SH07,LYMLASLK07}. The wave 
localization ensures 
the desired high $Q$-factors, whereas the unstable manifold provides the
directional emission.
Concerning the realisation of this scheme in practise,
adequate microcavity devices can be expected to be 
easy to fabricate as no sophisticated adjustment of parameters is required.
They are, moreover, well suited for multi-mode laser 
operation as all high-$Q$ modes of given polarization possess similar FFPs. 

\begin{figure}[ht]
\centerline{\includegraphics[width=1.0\figurewidth]{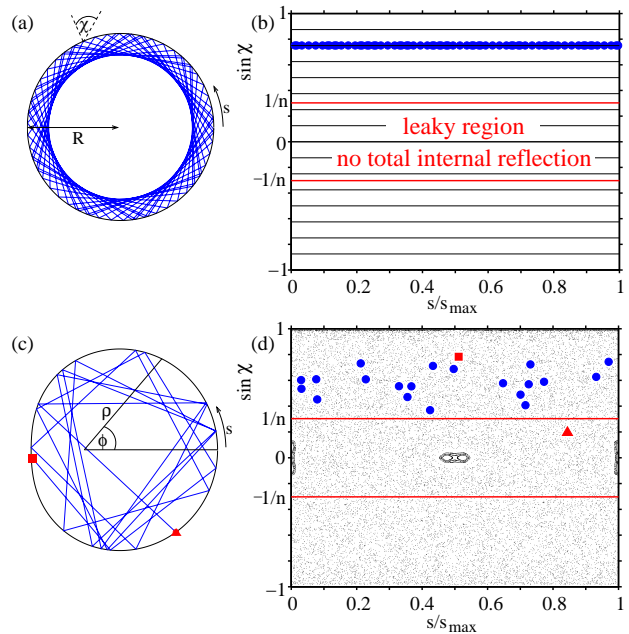}}
\caption{(color online). (a) Whispering-gallery ray trajectory in the circular
  microdisk. (b) Poincar\'e SOS 
  showing the trajectory (thick dots) in phase space; $s$ is the 
  arclength coordinate and $\chi$ is the angle of
  incidence. Typical trajectories fill a line of constant
  $\sin{\chi}$. The critical
  lines $\sin{\chi_c}=\pm 1/n$ enclose the leaky
  region.
(c) Chaotic ray trajectory in the {\limacon} cavity~(\ref{eq:robnik}) with
  $\varepsilon=0.43$.
(d) Poincar\'e SOS for the trajectory (thick dots) starting above the
  critical line (square) and refractively escaping after only 20
  bounces (triangle). Small dots show a typical trajectory in the
  corresponding closed billiard system.}
\label{fig:billiards}
\end{figure}

A microdisk is a quasi-two-dimensional geometry described by an effective index
of refraction~$n$. We assume $n=3.3$ (GaAs) both for transverse
magnetic (TM) and transverse electric (TE)  
polarization. A slight polarization dependence of $n$ is neglected;
it could be adjusted in the fabrication process, e.g., by changing
the slab thickness.
For the boundary curve of the deformed microdisk we choose the 
{\limacon} of Pascal which reads in polar coordinates~$(\rho,\phi)$ 
\begin{equation}\label{eq:robnik}
\rho(\phi) = R(1+\varepsilon\cos\phi) \ .
\end{equation}
The corresponding family of closed cavities is known
as \quoting{{\limacon} billiards}~\cite{Robnik83}. Ray and wave dynamics in
billiards has been 
extensively discussed in the fields of nonlinear 
dynamics~\cite{LichLieb92} and quantum chaos~\cite{Stoeckmann00}. 
The limiting case of vanishing deformation parameter $\varepsilon$ is the
circle with radius $R$. 
Figure~\ref{fig:billiards}(a) illustrates a whispering-gallery ray
trajectory in a circular microdisk trapped by total
internal reflection.  
A two-dimensional phase space representation, the so-called Poincar\'e surface of section
(SOS), is shown in \fig~\ref{fig:billiards}(b). Whenever the trajectory hits the
cavity's boundary, its position $s$ (arclength coordinate along the
circumference) and tangential momentum $\sin{\chi}$ (the angle of
incidence $\chi$ is measured from the surface normal) is recorded. 
For $\varepsilon=0$, rotational invariance of the system implies conservation
of the angular momentum~$\propto\sin{\chi}$. 
Ignoring wave effects, such a ray never leaves the cavity since it cannot enter
the leaky region between the two 
critical lines for total internal reflection given by $\sin{\chi_c}=\pm 1/n$. 

Figures~\ref{fig:billiards}(c) and (d) show a trajectory in the {\limacon}
cavity 
for $\varepsilon=0.43$. In contrast to the case of small deformation parameter
$\varepsilon$~\cite{Noeckel94} the dynamics is predominantly  
chaotic. Starting with an initial $\chi$ well above the critical line, a test
ray (square, thick dots, and triangle) rapidly 
approaches the leaky region ($\sin{\chi}$ is not conserved) 
where it escapes according to Snell's
and Fresnel's laws.   
Without refractive escape ($n=\infty$, hard wall or 
closed billiard limit), the trajectory would fill the phase
space in a random fashion (small dots).
Periodic ray trajectories do exist but
are always unstable, except for the two islands in the leaky region.
Whispering-gallery trajectories are confined to the tiny region $|\sin{\chi}|\apprge 0.99$.   

While in the long-time limit the phase space of closed chaotic systems is
essentially structureless, cf. \fig~\ref{fig:billiards}(d), the  phase space of
an open chaotic system is structured by the so-called \quoting{chaotic
  repeller}~\cite{LichLieb92}. It is the set of points in phase space that 
never visits the leaky  region both in forward and backward time evolution. 
The stable (unstable) manifold of
a chaotic repeller is the set of points that converges to the repeller in
forward (backward) time evolution. The unstable manifold therefore describes
the route of escape from the chaotic system. In the case of light, Fresnel's
laws impose an additional, polarization dependent weighting factor to the
unstable manifold in the leaky region~\cite{SH07}, since at each reflection the
intensity inside is multiplied by the Fresnel reflection coefficient.

\begin{figure}[ht]
\centerline{\includegraphics[width=1.0\figurewidth]{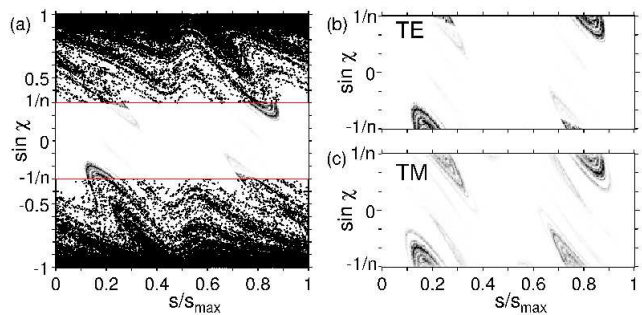}}
\caption{(color online). (a) Fresnel weighted unstable manifold of the {\limacon} cavity for TE
  polarization. Magnification reveals the differences between TE (b) and TM
  (c) polarization in the leaky region, which originate from Fresnel's 
  law. 
}
\label{fig:unstablemanifold}
\end{figure}

Following Refs.~\cite{LRKRK05,SH07} the unstable manifold can be uniquely
computed as a 
survival probability distribution calculated from an ensemble of rays starting
uniformly in phase space having identical intensity. 
Figure~\ref{fig:unstablemanifold} depicts the resulting Fresnel weighted 
unstable manifolds for the {\limacon} cavity using $50\,000$ rays. 
Note that (i) in the leaky region, the
manifold is concentrated on very few high-intensity 
spots. We therefore expect a highly directional FFP.
(ii) While in the case of TE polarization one
finds one spot with~$\chi>0$  (and another
symmetry-related one  at $s \rightarrow s_{\mbox{\footnotesize max}} -s, \chi
\rightarrow -\chi$), the TM polarization case possesses two of those. 

\begin{figure}[ht]
\centerline{\includegraphics[width=1.0\figurewidth]{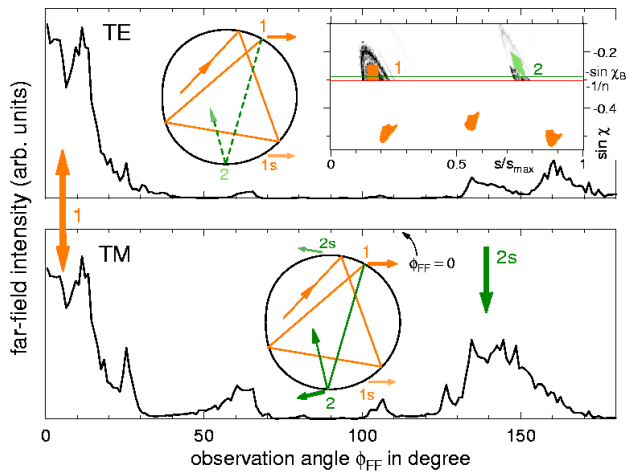}}
\caption{(color online). Far-field emission pattern for TE (top) and TM
polarization (bottom) calculated from an ensemble of rays on the unstable
manifold in \fig~\ref{fig:unstablemanifold}(b) and (c). 
Insets illustrate ray dynamics leading to directed emission. The dashed line
has strongly reduced intensity due to reflection near the Brewster angle. 
Right inset: 250~rays were started in the rectangular region at~1 and followed
forwards (region at~2) and backwards (the 3 regions below the critical
line). 1s and 2s mark the symmetry-related bounces to 1 and 2. 
} 
\label{fig:rayff}
\end{figure}

The unstable manifold in the leaky region directly determines the FFP. Mapping
the unstable manifold in Fig.~\ref{fig:unstablemanifold} to the far field by
using Snell's and Fresnel's laws (for generalization to curved interfaces, see~\cite{HS06}) we obtain Fig.~\ref{fig:rayff}.  
Note that the FFP is shown only for the upper half space
($0^\circ-180^\circ$), the lower half space is given by symmetry. 
For TE polarization, we find directionality around $\phi=0$, whereas in the
TM case additional, smaller peaks occur.
The ray in the left upper inset represents one typical trajectory emitting to 
$\phi \approx 0$ (marked by arrows). The emitting
bounce (marked 1, 1s is the symmetry-related counterpart), the three bounces
before and the one after (marked 2) are shown. 
Whereas the trajectories are equal for both polarizations, their intensities 
are different:
As visible in the right inset, the rays escaping at 1 hit
the line of the Brewster angle $|\chi_B|=\arctan 1/n<\arcsin 1/n$.  
In the TE case, transmission is nearly complete and no
intensity can reach the 
next bounce~2. This causes the sharp decrease in the intensity that 
is more clearly visible in \fig~\ref{fig:unstablemanifold}(b). 
%

The Fresnel law for TM polarization does not show the Brewster angle
feature. Therefore, a significant percentage of the light is reflected towards
bounce~2. Since bounce~2 emits into a different direction,  
an appreciable amount of intensity collects in a second far-field peak.

To summarize up to this point, we have seen that chaotic ray dynamics can lead
to highly directional emission. However, the $Q$-factor is low since the light
rays typically leave the cavity very quickly. Does this result of geometric
optics carry over to the wave dynamics of the electromagnetic field? 

It has been demonstrated that the FFP of optical modes can be strongly
influenced by the unstable manifold of the underlying ray 
dynamics~\cite{SRTCS04,LRKRK05,SH07}. A consequence is that for fixed
polarization and cavity parameters the FFP is independent on the internal mode
structure~\cite{LYMLASLK07}.   
This raises the hope that the high directionality observed in ray simulations
of the {\limacon} cavity will survive in wave optics.
To this end we solve Maxwell equations numerically using the boundary
element method~\cite{Wiersig02b}. According to the discrete symmetry, even and
odd modes are distinguished.

\begin{figure}[ht]
\centerline{\includegraphics[width=1.0\figurewidth]{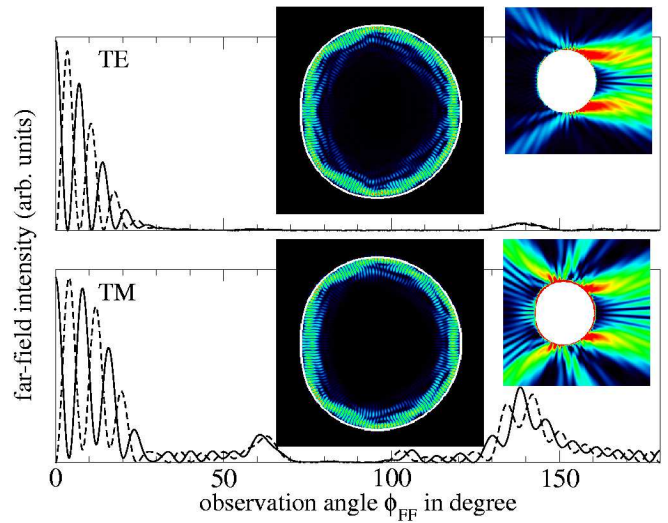}}
\caption{(color online). Angular dependence of the far-field electric field
intensity for a TE mode of even (top, solid line) and odd parity (top, dashed)
with $\Omega=26.0933$, $Q\approx 185\,000$, and a TM mode of even (bottom,
solid line) and odd parity (bottom, dashed) with $\Omega=25.8069$, $Q\approx
10^7$; cf.~\fig~\ref{fig:rayff}. Insets contain the near-field pattern of
even-parity modes (middle) and its external structure (right). 
}
\label{fig:modes}
\end{figure}

The top panel of \fig~\ref{fig:modes} shows near- and far-field pattern of a
high-$Q$ TE mode. The normalized frequency $\Omega=\omega R/c=26.0933$,
$c$ being the speed of light in vacuum, 
corresponds to, e.g., a free-space wavelength of about $900\,$nm for $R=3.75\,\mu$m. Indeed, as predicted by our ray dynamical analysis the mode exhibits 
directional light emission around $\phi=0$. The angular divergence of
$24^\circ$ is significantly smaller than the values reported for low-$Q$
disks~\cite{KLRK04,KTMJCC04}, and also less than in  
Ref.~\cite{WH06}. Moreover, for fixed polarization and cavity parameters, the
FFPs of all high-$Q$ TE modes in this cavity have similar envelope even though
the internal mode structure 
is in general different. This is exemplarily demonstrated in the
upper panel of \fig~\ref{fig:modes} for an odd-parity mode (dashed line)
which is quasi-degenerate with the even-parity solution (solid line). Note
that in other types of cavities the even- and the corresponding odd-parity
solution have in general a different FFP~\cite{WH06}.
For the high-$Q$ modes with TM polarization we also observe an FFP which is
independent of the internal mode structure, but a slightly 
different one, see lower panel of \fig~\ref{fig:modes}. 

Whereas the ray and wave based FFPs in \figs~\ref{fig:rayff} and
\ref{fig:modes}, respectively, agree remarkably well,  
other wave properties seem to contradict the ray simulations:  
(i) the mode does not look chaotic but spatially rather well confined.
(ii) the cavity $Q$s are too large if compared to the
escape rate from the chaotic repeller, in fact their values reach, or even
exceed, the present limit
achievable for microdisks 
with low residual absorption and surface roughness~\cite{MSJPLHKH07}.    

\begin{figure}[ht]
\centerline{\includegraphics[width=1.0\figurewidth]{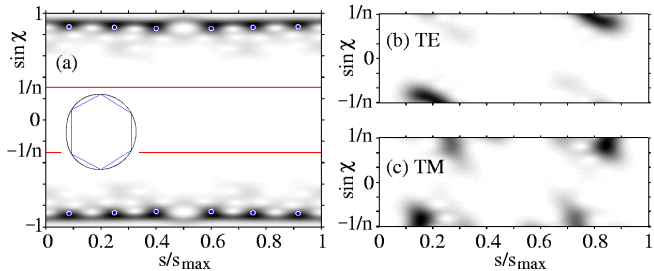}}
\caption{(color online). (a) Husimi projection of
 TE mode in \fig~\ref{fig:modes}, cf.  Poincar\'e SOS in
 \fig~\ref{fig:billiards}(d) and unstable manifold in
 \fig~\ref{fig:unstablemanifold}(a). The lines $\sin{\chi_c}=\pm 1/n$ enclosing
 the leaky region are indicated.  
 The dots mark the periodic ray trajectory illustrated in the inset. 
Magnified Husimi projection in the leaky region for the TE (b) and TM (c)
 mode; cf. \fig~\ref{fig:unstablemanifold}(b) and (c) for the ray simulation
 results. 
}
\label{fig:Husimi}
\end{figure}

To further investigate the character of these optical modes we consider the
Husimi projection~\cite{HSS03}, representing the wave analogue of the 
Poincar\'e SOS.  
From ray-wave correspondence one would expect that the Husimi projection
is distributed uniformly over the unstable manifold. However,
\fig~\ref{fig:Husimi}(a) demonstrates that the TE mode is localized around
$|\sin{\chi}|\approx 0.86$ and    
has only exponentially small intensity in the leaky region which 
explains the high $Q$-factor. A closer inspection~\cite{Hegger99} reveals that
the mode intensity is enhanced around an unstable periodic ray trajectory
(dots and inset in \fig~\ref{fig:Husimi}(a)) which is part of the chaotic
repeller. This phenomenon is called scarring~\cite{Heller84} and has been
observed in several kinds of physical systems including  
microcavities~\cite{LLCMKA02,FYC05,FHW06,Wiersig06}. 
Note that the other high-$Q$ modes found in this system also exhibit
localization along -- in general other -- unstable periodic rays.  
The Husimi projection of the TM mode in \fig~\ref{fig:modes} looks similar
(not shown). The localization is even stronger, leading to a higher
$Q$-factor.

Even though the Husimi projection has an exponentially small contribution in
the leaky region, it is precisely this outgoing light that determines the 
FFP. Figures~\ref{fig:Husimi}(b) and (c) show the Husimi projection
in the leaky region. The convincing agreement with the unstable manifold in
\figs~\ref{fig:unstablemanifold}(b) and (c) demonstrates its responsibility 
for the directional emission, whereas scarring
guarantees the high $Q$-factor.

Note that, due to the observed agreement between ray and wave simulation, our
results are also applicable to larger cavities.
The particular deformation parameter $\varepsilon = 0.43$ is the optimum 
value for the localization of the discussed FFPs with $n=3.3$ but highly
localized FFP and high $Q$-factors can also be found for
$0.41\apprle\varepsilon\apprle 0.49$ (not shown), i.e. fabrication  tolerances
are not crucial.  
Moreover, we tested that our results are robust against variations 
of the refractive index and remain valid for $2.7\apprle n\apprle 3.9$. 

In summary, we have proposed a deformed microdisk as a novel cavity
design for robust directional light emission from high-$Q$ modes. No
complicated adjustment of geometry parameters is necessary, and the emission
directionality is largely independent from wavelength, cavity size,
refractive index, and 
the details of the interior mode
structure. The latter finding is especially relevant for multi-mode lasing
devices. 
We trace our, at first sight, counterintuitive results back to (i) wave
localization along  
unstable periodic ray trajectories ensuring high $Q$-factors and (ii) escape
of rays along the unstable manifold of the chaotic repeller leading to
directional emission. 
The simplicity of the cavity design allows for easy fabrication with a wide
range of applications in photonics and optoelectronics. 
The discussed mechanisms are not restricted to disk-like geometries but can in
principle also be exploited for other geometries such as deformed microspheres
and microtori.  

We thank E.~Bogomolny, H.~Kantz, T.-Y.~Kwon, J.~Nagler, and M.~Robnik for 
discussions. Financial support by the Deutsche Forschungsgemeinschaft is
acknowledged.

\end{document}